\documentclass[aps,twocolumn,groupedaddress, showpacs,prb]{revtex4}
\usepackage{graphicx}%

\begin{document}
\bibliographystyle{apsrev}


\title{Microwave Bragg-scattering zone-axis-pattern analysis}


\author{P. Fraundorf}
\email[]{pfraundorf@umsl.edu}
\author{Bernard J. Feldman}
\author{W. Garver}
\author{M. Freeman}
\author{D. Proctor}
\affiliation{Physics \& Astronomy/Center for NanoScience, U. Missouri-StL (63121)}


\date{\today}

\begin{abstract}

Louis deBroglie's connection between momentum and spatial-frequency vectors is perhaps most viscerally-experienced via the real-time access that electron-diffraction provides to transverse slices of a nano-crystal's reciprocal-lattice. The classic introductory (and/or advanced) physics lab-experiment on microwave Bragg-scattering can with a bit of re-arrangement also give students access to ``zone-axis-pattern" slices through the 3D spatial-frequency (i.e. reciprocal) lattice of a ball-bearing crystal, which may likewise contain only a few unit-cells.

In this paper we show how data from the standard experimental set-up can be used to generate zone-axis-patterns oriented down the crystal rotation-axis. This may be used to give students direct experience with interpretation of lattice-fringe image power-spectra, and with nano-crystal electron-diffraction patterns, as well as with crystal shape-transforms that we use here to explain previously mis-identified peaks in the microwave data.

\end{abstract}
\maketitle

\tableofcontents
\section{Introduction}

Because of their high charge/mass ratio and hence strong interaction with matter, not to mention their wavelength in picometers, high-energy electrons are a benchmark tool for studying the interior of individual-structures on the nanoscale, at least to the extent that those structures will ``hold still" for a scattering experiment. Microwave scattering from a small ball-bearing lattice can be a way to give undergraduate students insight into the challenges of doing electron-scattering experiments on nanocrystalline materials.

Amato and Williams\cite{Amato2009} have previously discussed a way to modify classroom microwave-optics experiments\cite{Allen1955,Murray1974,Cornick2004,Yuan2011} to acquire ``X-ray powder diffraction" data on all Bragg peaks accessible from a two-dimensional lattice. In this paper, we discuss a way to put data from a standard set-up experiment into the format of an experimental electron zone-axis-pattern (ZAP). With this approach, students get some experience working with the crystal's reciprocal-lattice directly, and in the process gain some clues to the Fourier transform of a 3D crystal's shape\cite{Rees1950}. These effects of crystal shape become especially important when the crystal is only a few unit-cells across, in one or more directions.

\begin{figure}
\includegraphics[scale=.45]{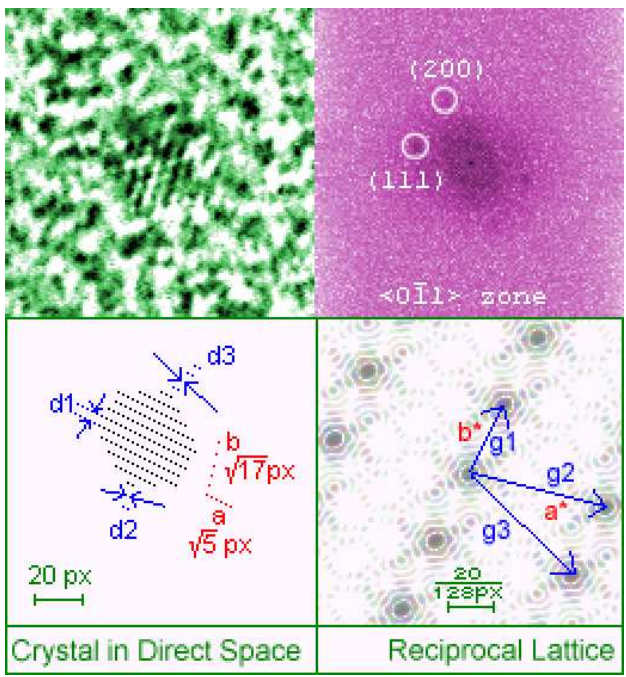}%
\caption{Top panels (a) show lattice-fringe image and power-spectrum of a Pt crystal in a 5.64 nm wide non-crystalline matrix field\cite{Mukherjee2013}, Bottom panels (b) show a 2D lattice with non-Cartesian basis-vectors and its Fourier-transform reciprocal-lattice, with three shape-transform convolved diffraction-spots (or g-vectors) labeled.}
\label{Fig1}
\end{figure}

In particular this experiment yields an experimental slice of our ball-bearing crystal's reciprocal-lattice, very much like electron-diffraction patterns\cite{Hirsch1965,Cowley1975,Williams1996,Fultz2001} and lattice-fringe image power-spectra\cite{Allpress1973,Spence1988,Fraundorf2005b,Wang2006,Kirkland2010} obtained from submicron-thick specimens in real-time (cf. Fig. \ref{Fig1}a). Our ``microwave slice" is perpendicular to the lattice (i.e. zone-axis) direction used as the crystal rotation-axis in the experiment. We further discuss how this window onto the distinct and complementary nature of direct/reciprocal dual vector-spaces can be enhanced by construction of a non-Cartesian ball-bearing lattice (cf. Fig. \ref{Fig1}b).

\section{Experimental procedures}

Our experimental system consisted of a $4 \times 4 \times 4$ simple-cubic ball-bearing lattice with lattice spacing $a \simeq b \simeq c \simeq 4.27$ cm and angles $\alpha \simeq \beta \simeq \gamma \simeq 90^o$ embedded into a styrofoam cube.  A Welch-system $10$-GHz klystron\cite{Marcley1960,Donnally1968,Bullen1969,Rossing1973} on one arm provides the microwave source, while a diode detector on the other arm measures the scattered microwave intensity.  A piece of aluminum foil is placed next to the styrofoam cube and between the source and detector arms to eliminate wavepaths not interior to the crystal.

\begin{figure}
\includegraphics[scale=0.65]{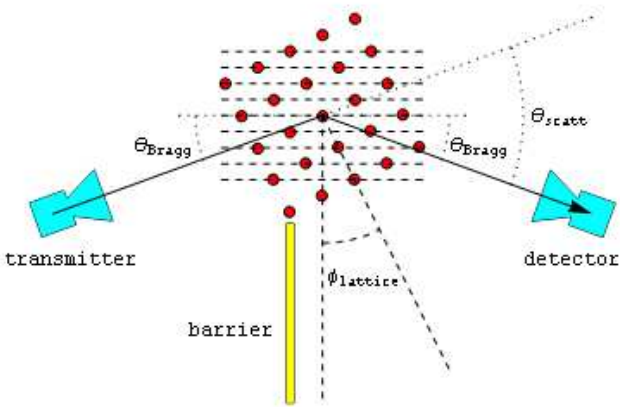}%
\caption{Angles for exploring $\{2n,n,0\}$-type reflections, with dashed horizontal-lines marking the (210) base-periodicity. Profiles over a range of Bragg-angles are obtained for each ``azimuthal" lattice-angle, which ``radial" scans are then projected onto a Cartesian field.}
\label{Fig1b}
\end{figure}

\begin{figure}
\includegraphics[scale=0.4]{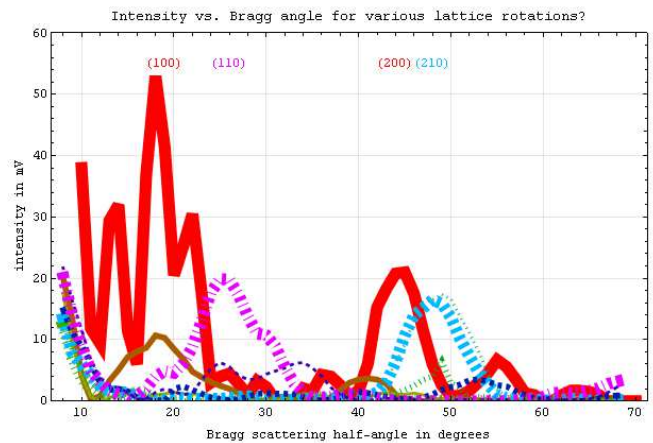}%
\caption{Intensity vs. Bragg angle for $\phi_{\text{lattice}}$ in degrees of 0, 5, 10 (solid, decreasing thickness), 15, 20, 25 (dotted), 30, 35, 40 (dashed), 45 (dot-dashed).}
\label{Fig2}
\end{figure}

The crystal orientation at an azimuthal angle $\phi_{\text{lattice}}$ is selected where $\phi_{\text{lattice}} = 0$ corresponds to the (100) lattice direction symmetric between the two arms of the system, as shown in Fig. \ref{Fig1b}. The scattered microwave intensity $I$ is measured at various grazing angles $\theta_{\text{Bragg}}$ between the source arm and the (vertical) ball-bearing plane of interest. The (horizontal) angle between the detector-arm and that ball-bearing plane is set to the same $\theta_{\text{Bragg}}$, so that intensity at a scattering angle of $2 \theta_{\text{Bragg}}$ is recorded. If the sample is a standard simple cubic ball bearing lattice, a quite complete data set can be obtained by recording such profiles for crystal orientations $\phi_{\text{lattice}}$ running from 0 to 45 degrees in 5 degree increments, as shown in Fig. \ref{Fig2}.

The wavelength of the microwave radiation was determined by removing the Styrofoam cube, setting $\theta_{\text{Bragg}}$ equal to 0 degrees, and varying the distance between the source and the detector in 1 mm steps. A sinusoidal variation in the intensity is observed due to creation of standing waves between the source and the detector. By fitting this data to a sinusoid term (and a linear term), a wavelength of $3.025 \pm 0.013$ cm was measured. This procedure is described in the Pasco manual\cite{Ayars2012}, and the result is consistent with the literature frequency.

\section{Data and analysis}

\begin{figure}
\includegraphics[scale=.75]{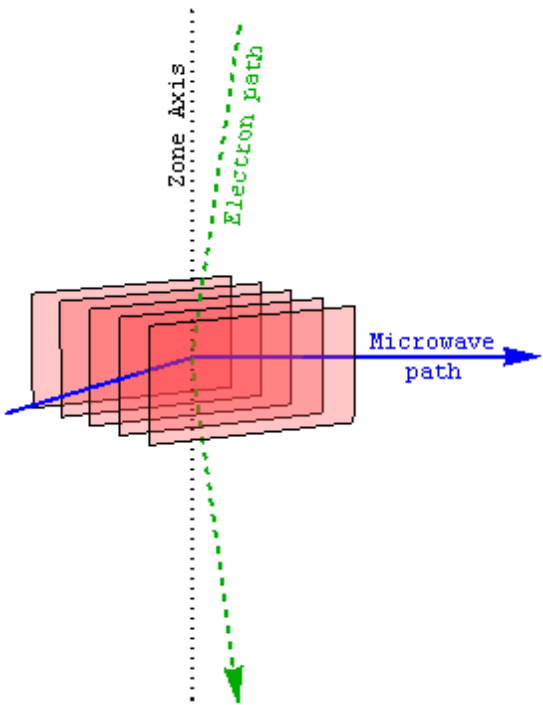}%
\caption{Comparison of electron (green dashed line) and microwave (blue line) zone-axis pattern geometries for a single lattice periodicity (red planes), with the rotation/zone-axis shown as a black dotted line. Zone axis patterns represent planar slices through the reciprocal lattice, which may be obtained either by diffraction or (via the Fourier-slice-theorem) by analysis of periodicities in the projected-potential down that same direction.}
\label{ZAPschematic}
\end{figure}

Even in non-cubic crystals the Bragg equation predicts momentum-changes $\Delta \vec{p} = h \vec{g} = \hbar \vec{k}$ of magnitude $2 \sin[\theta_{\text{Bragg}}] h/\lambda$ in reciprocal-lattice directions (hkl) normal to {\em planes of molecules} in a crystal's direct-space lattice. Lattice directions or zones [uvw] are similarly perpendicular to {\em planes of points} in the reciprocal-lattice. Thus a zone-axis-pattern is a map of projected scattering-power perpendicular to any lattice-direction in a periodic structure. In the Fraunhofer (far-field) diffraction-limit, such patterns are also the Fourier-transform of spatial-periodicities in the lattice projected down that direction i.e. 2D slices perpendicular to [uvw] through the crystal reciprocal (spatial-frequency) lattice \ref{ZAPschematic}. One can thus also think of zone-axis-patterns as diffraction patterns obtained using a flat (i.e. large-radius $1/\lambda$) Ewald-sphere.

In transmission electron microscopy (TEM), thanks to the small Bragg-angles (e.g. a quarter degree) and the small interaction mean-free-paths (requiring crystals well under a micron thick with elongate crystal shape-transforms\cite{Fraundorf2005b}), electron diffraction-patterns directly represent zone-axis patterns {\em down the direction of the beam}. These allow one to measure reciprocal-lattice periodicities, to form ``darkfield images" on active reflections (so-called g-vectors), and to set-up and interpret a wide range of other scattering experiments in real time\cite{Hirsch1965,Buxton1976,Spence1992,Reimer1997}.

Zone-axis-patterns further relate to direct-space lattice-images because zone-axis-patterns correspond to Fourier-transform power-spectra of ``projected-potential" lattice-images, via computer-aided-tomography's Fourier-slice-theorem in reverse i.e. the Fourier transform of an object's shadow represents a 2D slice through its frequency-space reciprocal-lattice. Of course only image power-spectra (as distinct from their complex Fourier transforms) are needed for comparison to the intensity-only information available from diffraction. For students taking modern physics, one might further note that electron phase-contrast lattice images connect to maps of projected-potential via a simple piecewise constant-potential proportionality to exit-surface deBroglie-phase, which microscope-optics turn into recordable intensity-variations in the wavefield downstream\cite{Spence1988}.

To construct a crystal rotation-axis zone-axis-pattern from our microwave data, we first re-parameterize the abcissa of the plots in Fig \ref{Fig2} to get intensity as a function of spatial-frequency $g$ instead of Bragg-angle $\theta_{\text{Bragg}}$, where from Bragg's Law the magnitude of the spatial frequency-vector $\vec{g}$ e.g. in [cycles/cm] is $g = 1/d = 2 \sin[\theta_{\text{Bragg}}]/\lambda$. If microwave wavelength is $\lambda \simeq 3$ cm, then Bragg-angle scans from 8 to 70 degrees examine spatial-frequency magnitudes $g$ ranging from about 0.09 to 0.62 [cycles/cm]. A ball-bearing periodicity of say $d_{100} \simeq 4.27$ cm will give us a peak at $g_{100} = 1/d_{100} \simeq 0.23$ [cycles/cm], and hence be easily detectable within this range.

\begin{figure*}
\includegraphics[scale=0.78]{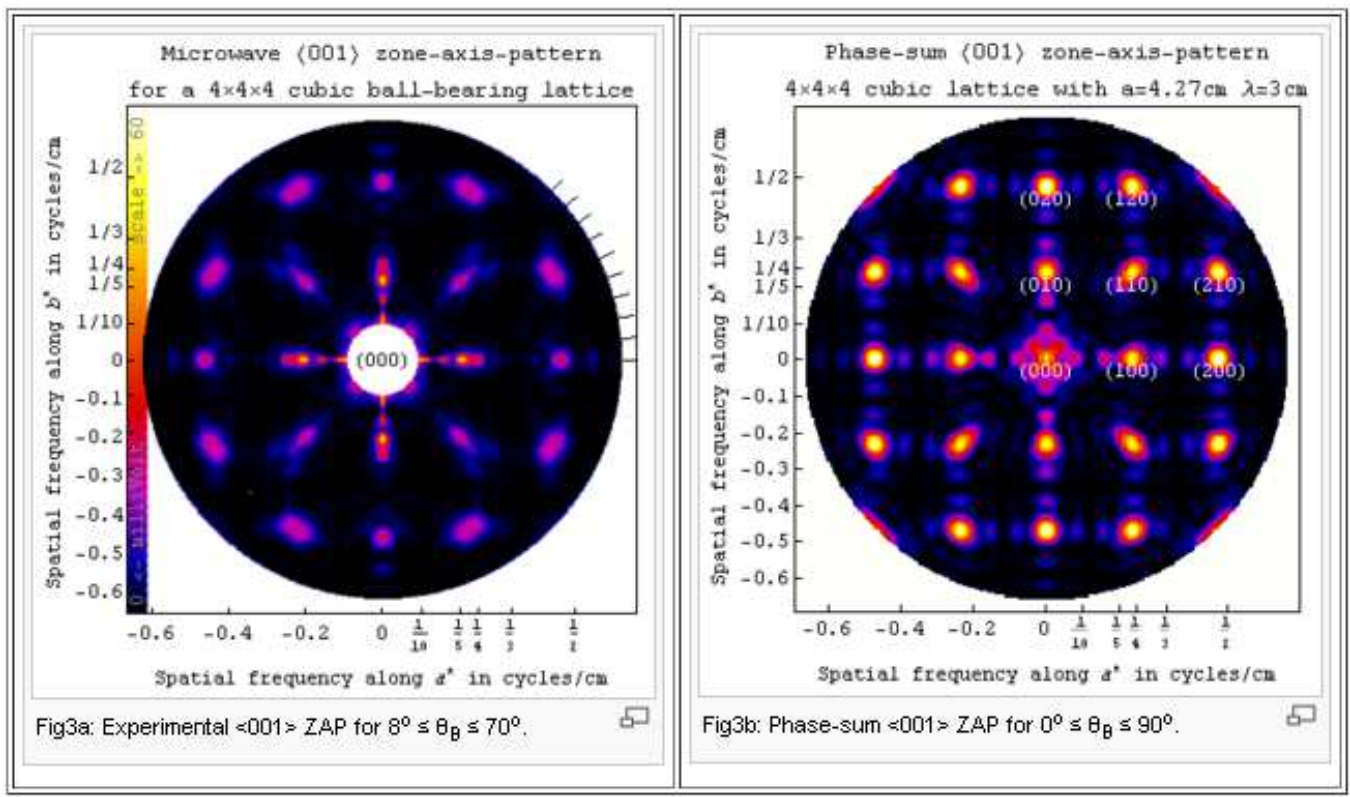}%
\caption{Comparison of (a) experimental and (b) phase-sum model zone-axis intensity-maps using $10$ GHz microwaves and a source/detector-lattice distance of 55cm.}
\label{Fig3}
\end{figure*}

Secondly, one then maps intensity as a function of spatial-frequency magnitude on a polar plot for the various possible crystal orientations flattice. If the simple-cubic lattice data has been taken for $\phi_{\text{lattice}}$ between 0 and 45 degrees from the $(h00)$ family of reflections, one can invoke the D4 (four-fold mirror) symmetry of a square in the rotation plane to fill in the pattern for values of $\phi_{\text{lattice}}$ from 0 to 360 degrees as shown in Fig. \ref{Fig3}a.

To compare one's experimental result with the scattering expected from the experimental arrangement of ball-bearings, a simple phase-sum model that focuses on the location (rather than the intensity) of zone-axis-pattern features is shown in Fig. \ref{Fig3}b. The model ignores intensity-variation with path-length and scattering-angle by just adding up the complex-phases for all scattering points to give an amplitude proportional to $\sum_{j} e^{\imath 2 \pi d_{j}/\lambda }$ where $d_{j}$ is the sum of source-to-scatterer and scatter-to-detector distances for the $j^{\text{th}}$ ball-bearing.. Each ball-bearing thus, for simplicity, contributes a unit-amplitude signal to the model sum. 

Sample code for using Mathematica to generate both experimental and model intensity maps is provided in the supplementary material for this paper\cite{ZAPsupplement2}. As you can see the phase-sum tells quite a bit about the location of reciprocal-lattice features in the zone-axis-pattern slice, although it would not be difficult for students to try predicting the effect of scattering-amplitudes on the pattern as well.

In both patterns, periodicities of the infinite crystal lattice show up as a square lattice of diffraction-spots or intensity-peaks. A standard set of reciprocal-lattice (Miller) indices for these peaks is provided in the positive quadrant of the model image.

Our finite crystal is truncated via multiplication in direct-space by a 3D window function that corresponds to its cubic shape. As a result the Fourier transform of this crystal shape function (i.e. the crystal's 3D shape transform) therefore convolves each of the points in the crystal's 3D reciprocal lattice.

For instance, a cube of side w has a shape transform that, in terms of the Cartesian components of spatial-frequency $\vec{g}$ e.g. in cycles/cm, looks like:
\begin{equation}
S[\vec{g}] = \frac{\sin[\pi w g_x]}{\pi g_x}\frac{\sin[\pi w g_y]}{\pi g_y}\frac{\sin[\pi w g_z]}{\pi g_z}
\end{equation}
where x, y and z are the $(100)$, $(010)$ and $(001)$ lattice directions for our faceted ball-bearing cube. This defines diffraction-peak broadening of half-width $1/w$ due to finite crystal size, as well as the $1/w$ periodicity of a series of damped ``sinc-oscillations" beginning at $1.5/w$ from peak center in a direction orthogonal to each crystal face. For a cube with $w = 4 \times d_{100} \simeq 17$ cm on a side, we therefore expect shape-transform peak half-widths and sinc-oscillation spacings in diffraction of $1/w \simeq 0.06$ [cycles/cm].

A zone-axis-pattern in the parallel-beam Fraunhofer (far-field) limit, as a planar slice through that reciprocal lattice, should therefore reveal around each diffraction spot a planar slice of the crystal's shape transform. Our phase-sum model, and our experimental ``divergent-beam diffraction-pattern", in addition contain Fresnel (near-field) diffraction effects although effects of both the infinite lattice (i.e. indexable diffraction-spots) and the shape-transform (in this case finite peak-widths and sinc-oscillations perpendicular to the cubic crystal facets) survive for source/detector-to-lattice distances more than 50 cm.

When intensities are taken into account e.g. by the experimental pattern, oscillations closest to the unscattered (central) beam-spot are easiest to see. In fact, the first (low-frequency side) sinc-oscillation associated with the (100) diffraction spot in Fig. \ref{Fig3} apparently shows up in the PASCO instruction manual data example\cite{Ayars2012}, even though it's incorrectly identified as ``a reflection off of a different plane than the one we're measuring".

\section{Experiment extensions}

Many yet unexplored threads are suggested by this analysis strategy. These include for example:

\begin{itemize}
  \item compare microwave zone-axis-patterns to the power spectrum of an image of your scattering-lattice projected down the pattern-direction,
  \item explore other shape-transform effects, like random-layering as occurs in some forms of graphite\cite{Warren1941} e.g. by random rotation of ball-bearing planes about the vertical rotation-axis,
  \item extend experiment \& modeling into the Fresnel near-field domain,
  \item construct and examine the dual vector-space of a non-Cartesian ball-bearing lattice,
  \item speed up and extend data acquisition with help from motorized and/or two-axis rotation, and
  \item construct a microwave intensity-model for more quantitative comparison to experimental data, perhaps taken with additional control of beam divergence/convergence.
\end{itemize}

\begin{figure}
\includegraphics[scale=0.36]{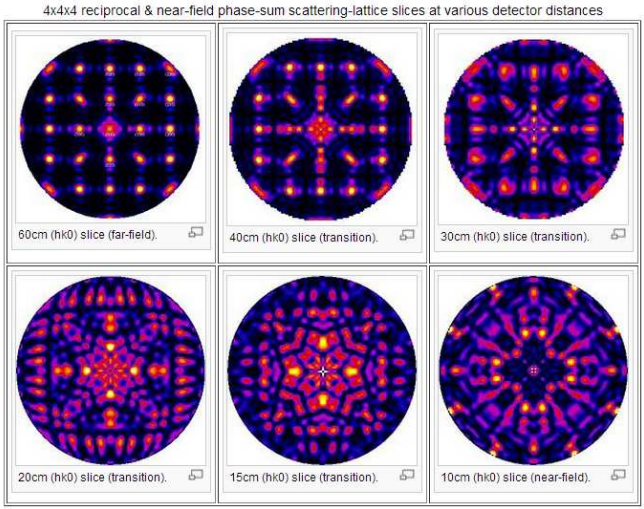}%
\caption{Phase-sum model for (commercially-available) $10.525$ GHz source/detector-to-crystal distances of 60, 50, 40 cm (top) and 30, 20, 10 cm (bottom).}
\label{Fig4}
\end{figure}

\begin{figure}
\includegraphics[scale=0.6]{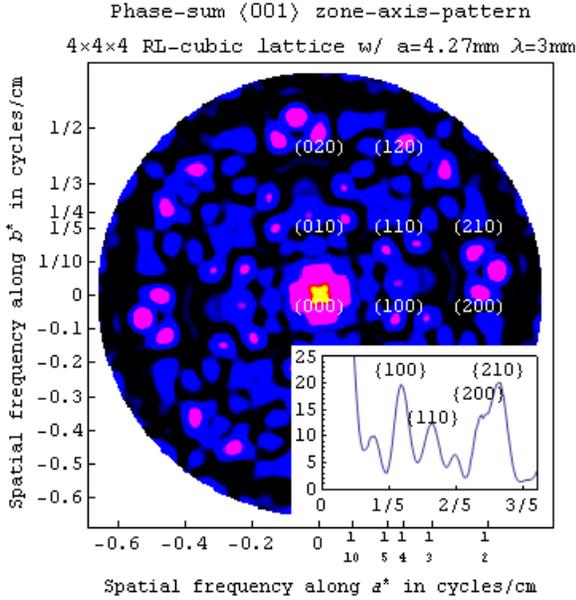}%
\caption{Phase-sum model for a $4\times4\times4$ ball-bearing lattice with ball-bearing sheet-rotations randomized \& contrast enhanced. The lower-right inset is an azimuthal average of intensities as a function of distance from the center.}
\label{RLLzap}
\end{figure}

Fig. \ref{Fig4} shows for example what the phase-sum model predicts for the pattern as one decreases source/detector to lattice distances. Fig. \ref{RLLzap} shows the spotty ``atom-thick sheet" powder-diffraction pattern that the phase-sum model predicts for the pattern if one rotates each of the four ball-bearing layers randomly about the rotation zone-axis. An azimuthal average of this pattern has \{001\}, \{110\}, \{200\} and \{210\} peaks whose breadth reflects the coherence-width of spacings in each sheet. To what extent these patterns can be matched by experiment remains to be seen.

The direction complementarity of reciprocal-lattice and direct-lattice vectors, with their co-variant as distinct from contra-variant transformation properties, is illustrated by a close look at Fig. \ref{Fig1}b. The basis vectors $\vec{a^*}$, $\vec{b^*}$ and $\vec{c^*}$ of the diffraction-spots in reciprocal space are not parallel to the direct-space basis vectors $\vec{a}$, $\vec{b}$ and $\vec{c}$, but are instead ``axial" vectors or one-forms perpendicular to those ``polar" direct-space vectors according to $\vec{a^*} = \vec{b} \times \vec{c}/V_c$, etc., where the unit cell volume is $V_c = \vec{a} \cdot (\vec{b} \times \vec{c} )$. Geologists are often more familiar than physicists with the elegant notation crystallographers have developed to deal with these dual vector-spaces, since minerals are much more likely than elemental solids to have low-symmetry lattices.

\begin{figure}
\includegraphics[scale=1.0]{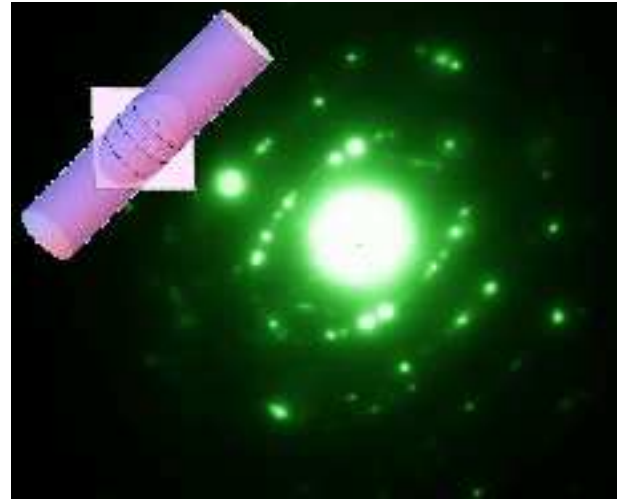}%
\caption{Random-layer-lattice hex-BN/C diffraction-pattern from a carbon-doped ZrB2 ceramic using 300 kV electrons, with a reciprocal-lattice model inset.}
\label{Fig7}
\end{figure}

For periodic lattices {\em projected into two dimensions}, only two basis vectors in frequency-space are needed to infer the rest of the 2D reciprocal unit-cell, and hence by Fourier-transformation the direct-space unit cell as well. Therefore a lattice with the angles shown in Fig. \ref{Fig1}b might have its lattice characterized by varying $\phi_{\text{lattice}}$ from 0 degrees clockwise by about 45 degrees to pick up the $g_2$ and $g_3$ spots from which the others (like $\vec{g}_1 = \vec{g}_2 - \vec{g}_3$) can be inferred. However the rotating-lattice technique of Amato and Williams\cite{Amato2009} would allow one to quickly scan all 360 degrees for a range of Bragg angles. Design of a two-axis eucentric goniometer would allow an even wider range of unknowns to be analyzed, although at this point the analysis might move beyond the ``hands on" scope of an advanced lab experiment.

Shape transforms have a breadth in frequency-space proportional to the inverse of their corresponding coherence-width (e.g. crystal size) in direct-space\cite{Hirsch1965}, as discussed in the previous section. In this context, the reciprocal-lattice of an atom-thick crystal is a spike (or ``rel-rod") in frequency space. A collection of parallel but randomly-rotated atomic layer-planes therefore has a cylindrical reciprocal-lattice, which can show up in the zone-axis-pattern as a circle when cut perpendicular to its axis, as parallel streaks when cut parallel to its axis, or as an oval\cite{Sasaki2001} like that shown in the experimental hexagonal-BN/C random-layer-lattice pattern in Fig. \ref{Fig7}. This effect might be explored with microwaves using a ball-bearing lattice by simply randomizing the azimuth of equally-spaced ball-bearing layers before taking the data.

Finally, the uncontrolled divergence of microwave intensity from the source and the variable direction-sensitivity of the detector further complicates the experimental data. Attempts to model these effects, and even better to control beam divergence with help from microwave optics upstream from the lattice, might do more than improve our quantitative understanding of the experimental data. Convergent beam electron diffraction is a case in point, in which an aperture-limited beam focused to a point on the specimen has opened up a new world of physics-based visualization to electron microscopists\cite{Buxton1976} including dispersion-surface profiles\cite{Chuvilin2005} plotted by the electrons themselves!

\begin{acknowledgments}
Matthew Freeman and David Proctor took the data for this project under the guidance of Bernard Feldman and Wayne Garver. Phil Fraundorf did the data conversions and the writeup. Thanks also to Bob Collins, Pat Sheehan, David Osborn, Greg Hilmas and Air Force contract number FA8650-05-D-5807 for help generating the data behind the oval diffraction pattern in Fig. \ref{Fig7}.
\end{acknowledgments}


\bibliography{ifzx2}

\begin{thebibliography}{28}
\expandafter\ifx\csname natexlab\endcsname\relax\def\natexlab#1{#1}\fi
\expandafter\ifx\csname bibnamefont\endcsname\relax
  \def\bibnamefont#1{#1}\fi
\expandafter\ifx\csname bibfnamefont\endcsname\relax
  \def\bibfnamefont#1{#1}\fi
\expandafter\ifx\csname citenamefont\endcsname\relax
  \def\citenamefont#1{#1}\fi
\expandafter\ifx\csname url\endcsname\relax
  \def\url#1{\texttt{#1}}\fi
\expandafter\ifx\csname urlprefix\endcsname\relax\def\urlprefix{URL }\fi
\providecommand{\bibinfo}[2]{#2}
\providecommand{\eprint}[2][]{\url{#2}}

\bibitem[{\citenamefont{Amato and Williams}(2009)}]{Amato2009}
\bibinfo{author}{\bibfnamefont{J.~C.} \bibnamefont{Amato}} \bibnamefont{and}
  \bibinfo{author}{\bibfnamefont{R.~E.} \bibnamefont{Williams}},
  \bibinfo{journal}{Am. J. Phys.} \textbf{\bibinfo{volume}{77}},
  \bibinfo{pages}{942} (\bibinfo{year}{2009}).

\bibitem[{\citenamefont{Allen}(1955)}]{Allen1955}
\bibinfo{author}{\bibfnamefont{R.~A.} \bibnamefont{Allen}},
  \bibinfo{journal}{Am. J. Phys.}
  \textbf{\bibinfo{volume}{23}}(\bibinfo{number}{5}), \bibinfo{pages}{297}
  (\bibinfo{year}{1955}).

\bibitem[{\citenamefont{Murray}(1974)}]{Murray1974}
\bibinfo{author}{\bibfnamefont{W.~H.} \bibnamefont{Murray}},
  \bibinfo{journal}{Am. J. Phys.} \textbf{\bibinfo{volume}{42}},
  \bibinfo{pages}{137} (\bibinfo{year}{1974}).

\bibitem[{\citenamefont{Cornick and Field}(2004)}]{Cornick2004}
\bibinfo{author}{\bibfnamefont{M.~T.} \bibnamefont{Cornick}} \bibnamefont{and}
  \bibinfo{author}{\bibfnamefont{S.~B.} \bibnamefont{Field}},
  \bibinfo{journal}{Am. J. Phys.} \textbf{\bibinfo{volume}{72}},
  \bibinfo{pages}{154} (\bibinfo{year}{2004}).

\bibitem[{\citenamefont{Yuan et~al.}(2011)\citenamefont{Yuan, Lin, Chang, and
  Shew}}]{Yuan2011}
\bibinfo{author}{\bibfnamefont{C.~P.} \bibnamefont{Yuan}},
  \bibinfo{author}{\bibfnamefont{S.~Y.} \bibnamefont{Lin}},
  \bibinfo{author}{\bibfnamefont{T.~H.} \bibnamefont{Chang}}, \bibnamefont{and}
  \bibinfo{author}{\bibfnamefont{B.~Y.} \bibnamefont{Shew}},
  \bibinfo{journal}{Am. J. Phys.} \textbf{\bibinfo{volume}{79}},
  \bibinfo{pages}{619} (\bibinfo{year}{2011}).

\bibitem[{\citenamefont{Rees and Spink}(1950)}]{Rees1950}
\bibinfo{author}{\bibfnamefont{A.~L.~G.} \bibnamefont{Rees}} \bibnamefont{and}
  \bibinfo{author}{\bibfnamefont{J.~A.} \bibnamefont{Spink}},
  \bibinfo{journal}{Acta Crystallographica} \textbf{\bibinfo{volume}{3}},
  \bibinfo{pages}{316} (\bibinfo{year}{1950}).

\bibitem[{\citenamefont{Hirsch1965}(1965/1977)}]{Hirsch1965}
\bibinfo{author}{\bibnamefont{Hirsch1965}}, \emph{\bibinfo{title}{Electron
  Microscopy of Thin Crystals}} (\bibinfo{publisher}{Krieger Publishing Co,
  Malabar FL}, \bibinfo{year}{1965/1977}).

\bibitem[{\citenamefont{Cowley}(1975)}]{Cowley1975}
\bibinfo{author}{\bibfnamefont{J.~M.} \bibnamefont{Cowley}},
  \emph{\bibinfo{title}{Diffraction Physics}}
  (\bibinfo{publisher}{North-Holland, Amsterdam}, \bibinfo{year}{1975}).

\bibitem[{\citenamefont{Williams and Carter}(1996,2009)}]{Williams1996}
\bibinfo{author}{\bibfnamefont{D.~B.} \bibnamefont{Williams}} \bibnamefont{and}
  \bibinfo{author}{\bibfnamefont{C.~B.} \bibnamefont{Carter}},
  \emph{\bibinfo{title}{Transmission Electron Microscopy}}
  (\bibinfo{publisher}{Springer Science, NY}, \bibinfo{year}{1996,2009}).

\bibitem[{\citenamefont{Fultz and Howe}(2001)}]{Fultz2001}
\bibinfo{author}{\bibfnamefont{B.}~\bibnamefont{Fultz}} \bibnamefont{and}
  \bibinfo{author}{\bibfnamefont{J.~M.} \bibnamefont{Howe}},
  \emph{\bibinfo{title}{Transmission electron microscopy and diffractometry of
  materials}} (\bibinfo{publisher}{Springer Science, NY},
  \bibinfo{year}{2001}).

\bibitem[{\citenamefont{Allpress and Sanders}(1973)}]{Allpress1973}
\bibinfo{author}{\bibfnamefont{J.~G.} \bibnamefont{Allpress}} \bibnamefont{and}
  \bibinfo{author}{\bibfnamefont{J.~V.} \bibnamefont{Sanders}},
  \bibinfo{journal}{J. Appl. Phys.} \textbf{\bibinfo{volume}{6}},
  \bibinfo{pages}{165} (\bibinfo{year}{1973}).

\bibitem[{\citenamefont{Spence}(1988)}]{Spence1988}
\bibinfo{author}{\bibfnamefont{J.~C.~H.} \bibnamefont{Spence}},
  \emph{\bibinfo{title}{Experimental high-resolution electron microscopy}}
  (\bibinfo{publisher}{Oxford University Press}, \bibinfo{year}{1988}),
  \bibinfo{edition}{2nd} ed.

\bibitem[{\citenamefont{Fraundorf et~al.}(2005)\citenamefont{Fraundorf, Qin,
  Moeck, and Mandell}}]{Fraundorf2005b}
\bibinfo{author}{\bibfnamefont{P.}~\bibnamefont{Fraundorf}},
  \bibinfo{author}{\bibfnamefont{W.}~\bibnamefont{Qin}},
  \bibinfo{author}{\bibfnamefont{P.}~\bibnamefont{Moeck}}, \bibnamefont{and}
  \bibinfo{author}{\bibfnamefont{E.}~\bibnamefont{Mandell}},
  \bibinfo{journal}{J. Appl. Phys.} \textbf{\bibinfo{volume}{98}},
  \bibinfo{pages}{114308} (\bibinfo{year}{2005}).

\bibitem[{\citenamefont{Wang et~al.}(2006)\citenamefont{Wang, Bleloch, Falke,
  and Goodhew}}]{Wang2006}
\bibinfo{author}{\bibfnamefont{P.}~\bibnamefont{Wang}},
  \bibinfo{author}{\bibfnamefont{A.~L.} \bibnamefont{Bleloch}},
  \bibinfo{author}{\bibfnamefont{U.}~\bibnamefont{Falke}}, \bibnamefont{and}
  \bibinfo{author}{\bibfnamefont{P.~J.} \bibnamefont{Goodhew}},
  \bibinfo{journal}{Ultramicroscopy} \textbf{\bibinfo{volume}{106}},
  \bibinfo{pages}{277} (\bibinfo{year}{2006}).

\bibitem[{\citenamefont{Kirkland}(2010)}]{Kirkland2010}
\bibinfo{author}{\bibfnamefont{E.~J.} \bibnamefont{Kirkland}},
  \emph{\bibinfo{title}{Advanced Computing in Electron Microscopy}}
  (\bibinfo{publisher}{Plenum Press, NY}, \bibinfo{year}{2010}),
  \bibinfo{edition}{2nd} ed.

\bibitem[{\citenamefont{Mukherjee et~al.}(2013)\citenamefont{Mukherjee,
  Ramalingam, Griggs, Hamm, Baker, Fraundorf, Sengupta, and
  Gangopadhyay}}]{Mukherjee2013}
\bibinfo{author}{\bibfnamefont{S.}~\bibnamefont{Mukherjee}},
  \bibinfo{author}{\bibfnamefont{B.}~\bibnamefont{Ramalingam}},
  \bibinfo{author}{\bibfnamefont{L.}~\bibnamefont{Griggs}},
  \bibinfo{author}{\bibfnamefont{S.}~\bibnamefont{Hamm}},
  \bibinfo{author}{\bibfnamefont{G.~A.} \bibnamefont{Baker}},
  \bibinfo{author}{\bibfnamefont{P.}~\bibnamefont{Fraundorf}},
  \bibinfo{author}{\bibfnamefont{S.}~\bibnamefont{Sengupta}}, \bibnamefont{and}
  \bibinfo{author}{\bibfnamefont{S.}~\bibnamefont{Gangopadhyay}},
  \bibinfo{journal}{Nanotechnology} \textbf{\bibinfo{volume}{23}},
  \bibinfo{pages}{485405} (\bibinfo{year}{2013}).

\bibitem[{\citenamefont{Marcley}(1960)}]{Marcley1960}
\bibinfo{author}{\bibfnamefont{R.~G.} \bibnamefont{Marcley}},
  \bibinfo{journal}{Am. J. Phys.} \textbf{\bibinfo{volume}{28}},
  \bibinfo{pages}{415} (\bibinfo{year}{1960}).

\bibitem[{\citenamefont{Donnally et~al.}(1968)\citenamefont{Donnally, Bradley,
  and Dewitt}}]{Donnally1968}
\bibinfo{author}{\bibfnamefont{B.~L.} \bibnamefont{Donnally}},
  \bibinfo{author}{\bibfnamefont{G.}~\bibnamefont{Bradley}}, \bibnamefont{and}
  \bibinfo{author}{\bibfnamefont{J.}~\bibnamefont{Dewitt}},
  \bibinfo{journal}{Am. J. Phys.} \textbf{\bibinfo{volume}{36}},
  \bibinfo{pages}{920} (\bibinfo{year}{1968}).

\bibitem[{\citenamefont{Bullen}(1969)}]{Bullen1969}
\bibinfo{author}{\bibfnamefont{T.~G.} \bibnamefont{Bullen}},
  \bibinfo{journal}{Am. J. Phys.} \textbf{\bibinfo{volume}{37}},
  \bibinfo{pages}{333} (\bibinfo{year}{1969}).

\bibitem[{\citenamefont{Rossing et~al.}(1973)\citenamefont{Rossing, Stadum, and
  Lang}}]{Rossing1973}
\bibinfo{author}{\bibfnamefont{T.~D.} \bibnamefont{Rossing}},
  \bibinfo{author}{\bibfnamefont{R.}~\bibnamefont{Stadum}}, \bibnamefont{and}
  \bibinfo{author}{\bibfnamefont{D.}~\bibnamefont{Lang}}, \bibinfo{journal}{Am.
  J. Phys.} \textbf{\bibinfo{volume}{41}}, \bibinfo{pages}{129}
  (\bibinfo{year}{1973}).

\bibitem[{\citenamefont{Ayars}(2012)}]{Ayars2012}
\bibinfo{author}{\bibfnamefont{E.}~\bibnamefont{Ayars}}, in
  \emph{\bibinfo{booktitle}{Instruction Manual and Experiment Guide for the
  PASCO scientifiic Model WA-9314B: Microwave Optics}}
  (\bibinfo{publisher}{PASCO scientific}, \bibinfo{year}{2012}),
  \bibinfo{number}{012-04630G}, pp. \bibinfo{pages}{35--44}.

\bibitem[{\citenamefont{Buxton et~al.}(1976)\citenamefont{Buxton, Eades,
  Steeds, and Rackham}}]{Buxton1976}
\bibinfo{author}{\bibfnamefont{B.~F.} \bibnamefont{Buxton}},
  \bibinfo{author}{\bibfnamefont{J.~A.} \bibnamefont{Eades}},
  \bibinfo{author}{\bibfnamefont{J.~W.} \bibnamefont{Steeds}},
  \bibnamefont{and} \bibinfo{author}{\bibfnamefont{R.~M.}
  \bibnamefont{Rackham}}, \bibinfo{journal}{Philosophical Transactions of the
  Royal Society of London, Series A} \textbf{\bibinfo{volume}{281}},
  \bibinfo{pages}{171} (\bibinfo{year}{1976}).

\bibitem[{\citenamefont{Spence and Zuo}(1992)}]{Spence1992}
\bibinfo{author}{\bibfnamefont{J.~C.~H.} \bibnamefont{Spence}}
  \bibnamefont{and} \bibinfo{author}{\bibfnamefont{J.}~\bibnamefont{Zuo}},
  \emph{\bibinfo{title}{Electron microdiffraction}} (\bibinfo{publisher}{Plenum
  Press, NY}, \bibinfo{year}{1992}).

\bibitem[{\citenamefont{Reimer}(1997)}]{Reimer1997}
\bibinfo{author}{\bibfnamefont{L.}~\bibnamefont{Reimer}},
  \emph{\bibinfo{title}{Transmission electron microscopy: Physics of image
  formation and microanalysis}} (\bibinfo{publisher}{Springer Science, NY},
  \bibinfo{year}{1997}), \bibinfo{edition}{4th} ed.

\bibitem[{\citenamefont{Fraundorf}(2013)}]{ZAPsupplement2}
\bibinfo{author}{\bibfnamefont{P.}~\bibnamefont{Fraundorf}},
  \emph{\bibinfo{title}{Sample code for microwave bragg-scattering
  zone-axis-pattern analysis}} (\bibinfo{year}{2013}),
  \urlprefix\url{http://tinyurl.com/m35pfbn}.

\bibitem[{\citenamefont{Warren}(1941)}]{Warren1941}
\bibinfo{author}{\bibfnamefont{B.~E.} \bibnamefont{Warren}},
  \bibinfo{journal}{Phys. Rev.} \textbf{\bibinfo{volume}{59}},
  \bibinfo{pages}{963} (\bibinfo{year}{1941}).

\bibitem[{\citenamefont{Sasaki et~al.}(2001)\citenamefont{Sasaki, Ebina,
  Kitami, and Watanabe}}]{Sasaki2001}
\bibinfo{author}{\bibfnamefont{T.}~\bibnamefont{Sasaki}},
  \bibinfo{author}{\bibfnamefont{Y.}~\bibnamefont{Ebina}},
  \bibinfo{author}{\bibfnamefont{Y.}~\bibnamefont{Kitami}}, \bibnamefont{and}
  \bibinfo{author}{\bibfnamefont{M.}~\bibnamefont{Watanabe}},
  \bibinfo{journal}{J. Phys. Chem. B} \textbf{\bibinfo{volume}{105}},
  \bibinfo{pages}{6116} (\bibinfo{year}{2001}).

\bibitem[{\citenamefont{Chuvilin and Kaiser}(2005)}]{Chuvilin2005}
\bibinfo{author}{\bibfnamefont{A.}~\bibnamefont{Chuvilin}} \bibnamefont{and}
  \bibinfo{author}{\bibfnamefont{U.}~\bibnamefont{Kaiser}},
  \bibinfo{journal}{Ultramicroscopy} \textbf{\bibinfo{volume}{104}},
  \bibinfo{pages}{73} (\bibinfo{year}{2005}).

\end{thebibliography}


\end{document}